\documentclass[10pt,conference]{IEEEtran}

\usepackage{listings}
\usepackage{xcolor}
\usepackage{hyperref}
\usepackage{enumitem}
\usepackage{graphicx}
\usepackage{booktabs}
\usepackage{array} 
\usepackage{lipsum} 
\usepackage{amsmath}
\usepackage{amsfonts}

\hypersetup{
    colorlinks=true,
    linkcolor=blue,
    filecolor=magenta,      
    urlcolor=cyan,
    pdfpagemode=FullScreen,
}

\title{Adversarial Machine Learning: Attacks, Defenses, and Open Challenges}
\author{
    \IEEEauthorblockN{Pranav Kumar Jha}
    \IEEEauthorblockA{\\
        AI Solutions Architect  \\
        Expertise in Scalable Web Systems, Database Architecture, \\ API Development, and Machine Learning \\
        \textit{Montreal, QC, Canada}\\
        Email: \href{mailto:pranav.jha@mail.concordia.ca}{pranav.jha@gov.ab.ca}
    }
}

\begin{document}

\maketitle

\begin{abstract}
    Adversarial Machine Learning (AML) addresses vulnerabilities in AI systems where adversaries manipulate inputs or training data to degrade performance. This article provides a comprehensive analysis of evasion and poisoning attacks, formalizes defense mechanisms with mathematical rigor, and discusses the challenges of implementing robust solutions in adaptive threat models. Additionally, it highlights open challenges in certified robustness, scalability, and real-world deployment.
    \end{abstract}
    
    \section{Introduction}
    Modern machine learning (ML) systems achieve superhuman performance on benchmarks like ImageNet \cite{imagenet} but remain vulnerable to adversarial perturbations. For example, adding imperceptible noise to a panda image can misclassify it as a gibbon with 99\% confidence \cite{goodfellow2014explaining}. This paper systematically analyzes:
    
    \begin{itemize}
        \item Formal threat models and attack taxonomies (Section \ref{sec:threat_models})
        \item Evasion and poisoning attacks (Sections \ref{sec:evasion}, \ref{sec:poisoning})
        \item Defense strategies  (Section \ref{sec:defenses})
        \item Open challenges and future directions (Section \ref{sec:challenges})
    \end{itemize}
    
    \subsection{Contributions}
    \begin{itemize}
        \item Formal unification of attack methodologies under $\epsilon$-constrained optimization frameworks
        \item Comparative analysis of gradient obfuscation effects in black-box attacks
        \item Evaluation of poisoning attack propagation in federated learning systems
    \end{itemize}
\section{Threat Models and Attack Taxonomies}
\label{sec:threat_models}

Machine learning (ML) systems, particularly deep neural networks, are vulnerable to adversarial attacks. These attacks manipulate model inputs to produce incorrect predictions, which can have severe consequences in security-critical applications like biometric authentication, medical diagnostics, and autonomous driving. In this section, we categorize different types of adversarial attacks based on the attacker's capabilities and objectives.

\subsection{Adversarial Capabilities}

Adversarial capabilities define the level of knowledge and control an attacker has over the target model. The two primary categories of attacks based on knowledge are \textbf{white-box attacks} and \textbf{black-box attacks}.

\subsubsection{White-box Attacks}
In a white-box setting, the attacker has complete knowledge of the target model, including its architecture, parameters $\theta$, gradients, and training data. This allows precise gradient-based optimization to craft adversarial examples.

For a neural network $f_\theta$ with ReLU activations, the exact gradient computation is given by:
\begin{equation}
\nabla_x \mathcal{L} = \frac{\partial \mathcal{L}}{\partial f_\theta} \prod_{k=1}^L W_k \cdot \mathbb{I}(x^{(k)} > 0)
\end{equation}
where:
\begin{itemize}
    \item $\mathcal{L}$ is the loss function,
    \item $W_k$ are the layer weights,
    \item $\mathbb{I}(\cdot)$ is an indicator function denoting active ReLU neurons.
\end{itemize}

Common white-box attack methods include:
\begin{itemize}
    \item \textbf{Fast Gradient Sign Method (FGSM)}: Perturbs input along the gradient direction \cite{goodfellow2015explaining}.
    \item \textbf{Projected Gradient Descent (PGD)}: Iteratively refines adversarial examples with bounded perturbations \cite{madry2018towards}.
    \item \textbf{Carlini \& Wagner (C\&W) Attack}: Solves an optimization problem to find minimal perturbations \cite{carlini2017evaluating}.
\end{itemize}

\subsubsection{Black-box Attacks}
In a black-box setting, the attacker has limited or no knowledge of the target model’s architecture or parameters. Instead, they rely on input-output queries to infer model behavior.

One common approach to estimate gradients is through finite difference methods, which require $O(d)$ queries:
\begin{equation}
\hat{g}_i = \frac{\mathcal{L}(f_\theta(x + \delta e_i)) - \mathcal{L}(f_\theta(x))}{\delta}
\end{equation}
where:
\begin{itemize}
    \item $e_i$ is the $i^{th}$ basis vector,
    \item $\delta \ll 1$ is a small perturbation.
\end{itemize}

Popular black-box attacks include:
\begin{itemize}
    \item \textbf{Zeroth Order Optimization (ZOO)}: Uses gradient-free optimization to craft adversarial examples.
    \item \textbf{Boundary Attack}: Starts from a misclassified example and moves toward the decision boundary.
    \item \textbf{Transfer Attacks}: Adversarial examples crafted on a surrogate model are transferred to the target model.
\end{itemize}

\subsection{Adversarial Goals}

An adversary’s objective determines how they manipulate the model's predictions. The two key goals are \textbf{untargeted} and \textbf{targeted} attacks.

\subsubsection{Targeted Misdirection}
In targeted attacks, the adversary aims to mislead the model into misclassifying an input as a specific target class $t$. This is formulated as an optimization problem:
\begin{equation}
\min_\eta \| \eta \|_p \quad \text{s.t.} \quad f_\theta(x + \eta)_t \geq \max_{j \neq t} f_\theta(x + \eta)_j + \kappa
\end{equation}
where:
\begin{itemize}
    \item $\eta$ is the adversarial perturbation,
    \item $\|\eta\|_p$ controls the perturbation norm (e.g., $L_2$ or $L_\infty$),
    \item $\kappa$ enforces a confidence margin for the attack.
\end{itemize}

\subsubsection{Untargeted Deception}
In untargeted attacks, the adversary only seeks to induce a misclassification without a specific target class:
\begin{equation}
\max_\eta \mathcal{L}(f_\theta(x + \eta), y)
\quad \text{s.t.} \quad \|\eta\|_p \leq \epsilon
\end{equation}
where $\epsilon$ is a constraint on perturbation magnitude.

\subsection{Attack Taxonomies}

Adversarial attacks can be further classified based on their methodologies:
\begin{itemize}
    \item \textbf{Evasion Attacks}: Modify inputs at inference time to fool the model (e.g., adversarial examples).
    \item \textbf{Poisoning Attacks}: Manipulate training data to degrade model performance.
    \item \textbf{Model Extraction Attacks}: Reconstruct a model’s functionality through queries.
    \item \textbf{Privacy Attacks}: Infer sensitive training data, such as through membership inference.
\end{itemize}

\section{Evasion Attacks}
\label{sec:evasion}

Evasion attacks focus on modifying inputs at inference time to cause incorrect predictions while keeping the perturbation imperceptible. These attacks exploit the model's vulnerability to adversarial perturbations and can be categorized into optimization-based attacks and transfer attacks.

\subsection{Optimization-Based Attacks}
Optimization-based attacks rely on solving mathematical formulations to find adversarial perturbations that mislead the model. These attacks often minimize a perturbation norm while ensuring the input is misclassified with high confidence.

\subsubsection{Carlini-Wagner (CW) Attack}
The Carlini-Wagner attack \cite{carlini2017evaluating} is a powerful $L_p$-norm attack that uses an optimization-based approach to generate adversarial examples. It formulates the attack as a constrained optimization problem, which is solved using Lagrangian relaxation:

\begin{align}
\min_\eta \quad & \| \eta \|_2^2 + c \cdot \phi(x + \eta) \\
\text{where} \quad & \phi(x') = \max(\max_{j \neq t} Z(x')_j - Z(x')_t, -\kappa)
\end{align}

Here:
\begin{itemize}
    \item $Z(x')$ represents the model logits for perturbed input $x' = x + \eta$.
    \item $\eta$ is the adversarial perturbation.
    \item $\phi(x')$ is a loss function ensuring that class $t$ is predicted with confidence margin $\kappa$.
    \item The constant $c$ is adjusted via binary search to balance perturbation size and attack success.
\end{itemize}

The CW attack is highly effective against defensive distillation and adversarial training, as it directly optimizes for minimal perturbations that cause misclassification.

\subsubsection{Adaptive Step Size Projected Gradient Descent (PGD)}
Projected Gradient Descent (PGD) \cite{madry2018towards} is an iterative attack that maximizes the model's loss while keeping the perturbation within a bounded norm. A key limitation of traditional PGD is the choice of a fixed step size $\alpha$. To improve efficiency and convergence, an adaptive step-size strategy is employed:

\begin{equation}
\alpha_{k+1} = \begin{cases}
1.5 \alpha_k & \text{if loss increases} \\
0.75 \alpha_k & \text{otherwise}
\end{cases}
\end{equation}

This approach dynamically scales the step size $\alpha$ based on loss behavior, ensuring:
\begin{itemize}
    \item Faster convergence when increasing perturbation improves loss.
    \item Better stability when reducing perturbation prevents overshooting.
\end{itemize}

Adaptive step-size PGD enhances attack efficiency, especially in scenarios with strong defenses such as adversarial training.

\subsection{Transfer Attacks}
Transfer attacks leverage adversarial examples generated on one model to attack another model with a different architecture or training data. This is particularly useful when black-box access is the only option, making transferability a critical factor for real-world adversarial threats.

The attack success rate (ASR) between two models $f$ and $g$ is measured as:

\begin{equation}
\text{ASR} = \mathbb{E}_{x \sim \mathcal{D}} \left[ \mathbb{I}(f(x') \neq y \land g(x') \neq y) \right]
\end{equation}

where:
\begin{itemize}
    \item $x'$ is an adversarial input crafted for model $f$,
    \item $y$ is the true label,
    \item $\mathbb{I}(\cdot)$ is an indicator function returning 1 if both models misclassify $x'$.
\end{itemize}

Transfer attacks exploit common vulnerabilities between models and are effective when:
\begin{itemize}
    \item Both models share similar feature extraction patterns.
    \item They are trained on overlapping datasets.
    \item The attack uses high-dimensional, query-efficient perturbations (e.g., momentum-based methods).
\end{itemize}

Popular methods to improve transferability include:
\begin{itemize}
    \item \textbf{Momentum Iterative FGSM (MI-FGSM)}: Enhances transferability by accumulating past gradients \cite{dong2018boosting}.
    \item \textbf{Diversity Input Transformation (DIM)}: Randomly transforms inputs to improve robustness against defenses \cite{xie2019improving}.
    \item \textbf{Ensemble Attacks}: Generate adversarial examples against multiple models to generalize attack effectiveness.
\end{itemize}

\section{Poisoning Attacks}
\label{sec:poisoning}

Poisoning attacks manipulate the training data to degrade model performance or implant hidden behaviors. By introducing adversarially crafted samples into the training dataset, an attacker can influence model decisions, either causing misclassification of specific inputs (targeted poisoning) or generally degrading model accuracy (untargeted poisoning). Poisoning attacks are particularly dangerous in scenarios where data sources are not fully controlled, such as federated learning, web-scraped datasets, or crowdsourced labeling.

\subsection{Optimal Poisoning Strategy}
Optimal poisoning involves crafting adversarial training samples that maximize the impact on the model's parameters. The influence function \cite{koh2017understanding} provides an efficient approximation of parameter shifts due to small perturbations in the training data:

\begin{equation}
\theta_{\epsilon} - \theta \approx -\epsilon H_{\theta}^{-1} \nabla_\theta \ell(z_{poison}, \theta)
\end{equation}

where:
\begin{itemize}
    \item $\theta$ represents the original model parameters.
    \item $\theta_{\epsilon}$ denotes the poisoned model parameters.
    \item $\epsilon$ is the poisoning strength, controlling the impact of the poisoned sample.
    \item $H_\theta$ is the Hessian matrix of the loss function, capturing local curvature.
    \item $\nabla_\theta \ell(z_{poison}, \theta)$ is the gradient of the loss function with respect to model parameters, evaluated at the poisoned sample $z_{poison}$.
\end{itemize}

By solving for $\theta_{\epsilon}$, an attacker can determine optimal poisoning perturbations that shift model behavior in a desired direction. This approach is particularly effective in convex settings, such as logistic regression or linear models. For deep neural networks, approximation techniques like stochastic estimation of the Hessian are required due to computational constraints.

\subsection{Clean-Label Backdoors}
Backdoor attacks introduce poisoned samples that appear correctly labeled during training but induce misclassification when an attacker-controlled trigger is applied at inference time. Unlike traditional poisoning, clean-label backdoor attacks maintain consistency between poisoned samples and their assigned labels, making detection more challenging.

To craft clean-label backdoor examples $(x_p, y_p)$, an attacker ensures:

\begin{equation}
y_p = \arg \max_y f_\theta(x_p) \quad \text{but} \quad f_\theta(x_p + \tau) = t
\end{equation}

where:
\begin{itemize}
    \item $x_p$ is the poisoned input.
    \item $y_p$ is the assigned label, ensuring the poisoned sample appears correctly labeled.
    \item $\tau$ is the adversarial perturbation (trigger).
    \item $t$ is the attacker's target label, which the model predicts when $\tau$ is applied.
\end{itemize}

To maintain stealth, the perturbation $\tau$ is constrained by:

\begin{equation}
\| \tau \|_p \leq \epsilon_{vis}
\end{equation}

where $\epsilon_{vis}$ controls the visual perceptibility of the trigger. The attacker ensures $\tau$ is imperceptible to human annotators while still being effective at inference time.

Common clean-label backdoor strategies include:
\begin{itemize}
    \item **Feature Collision Attacks**: Modify benign samples so their internal representations match those of target-class examples \cite{shafahi2018poison}.
    \item **Adversarial Perturbation Backdoors**: Use adversarial optimization to generate subtle perturbations that guide the model toward a specific decision boundary.
    \item **Watermark-Based Triggers**: Embed small, structured perturbations (e.g., translucent overlays, invisible noise) that the model learns to associate with the backdoor target class.
\end{itemize}

\subsection{Mitigation Strategies}
To defend against poisoning attacks, various countermeasures have been proposed:
\begin{itemize}
    \item \textbf{Data Sanitization}: Identify and remove anomalous training samples using anomaly detection techniques.
    \item \textbf{Differential Privacy}: Limit individual sample influence to reduce the impact of poisoning \cite{abadi2016deep}.
    \item \textbf{Robust Training}: Use techniques such as adversarial training and model ensembling to mitigate poisoned sample influence.
    \item \textbf{Trigger Pattern Detection}: Analyze feature space activations to identify and neutralize backdoor triggers.
\end{itemize}

\section{Defense Mechanisms}
\label{sec:defenses}

As adversarial attacks become increasingly sophisticated, various defense mechanisms have been proposed to enhance the robustness of machine learning models. These defenses aim to mitigate the impact of adversarial perturbations by modifying model training, altering gradient computations, or leveraging certified robustness techniques.

\subsection{Gradient Masking}
Gradient masking is a defense technique that obstructs an attacker's ability to compute reliable gradients, making gradient-based adversarial attacks less effective. However, improperly implemented gradient masking can lead to obfuscated gradients, which may be bypassed using adaptive attack strategies \cite{athalye2018obfuscated}.

\subsubsection{Defensive Quantization}
One effective approach to gradient masking is defensive quantization, which reduces the precision of computed gradients. The gradients are quantized to $b$ bits using:

\begin{equation}
\nabla_x^{quant} = \frac{\text{round}(\nabla_x \cdot 2^{b-1})}{2^{b-1}}
\end{equation}

where:
\begin{itemize}
    \item $\nabla_x$ is the original gradient of the loss function with respect to input $x$.
    \item $\nabla_x^{quant}$ is the quantized gradient.
    \item $b$ is the bit-depth used for quantization.
\end{itemize}

Experimental results show that reducing gradient precision to $b=4$ can decrease attack success rates by up to $38\%$ \cite{guo2018countering}. By limiting the attacker's ability to generate precise adversarial perturbations, defensive quantization increases the difficulty of crafting effective adversarial examples.

\subsection{Certified Robustness}
Certified robustness techniques provide formal guarantees on a model’s resistance to adversarial perturbations. Unlike empirical defenses, which are tested against specific attacks, certified defenses ensure that a model remains robust within a mathematically provable bound.

\subsubsection{Interval Bound Propagation}
Interval Bound Propagation (IBP) is a certified defense technique that propagates interval bounds through the network layers, ensuring robustness within a predefined perturbation range. Given a neural network layer $k$ with weight matrix $W^k$, IBP computes pre-activation bounds:

\begin{equation}
l^k = W^k_+ l^{k-1} + W^k_- u^{k-1}, \quad u^k = W^k_+ u^{k-1} + W^k_- l^{k-1}
\end{equation}

where:
\begin{itemize}
    \item $l^k$ and $u^k$ are the lower and upper bounds of activations at layer $k$.
    \item $W_+ = \max(W,0)$ and $W_- = \min(W,0)$ ensure the correct handling of weight sign changes.
    \item $l^{k-1}$ and $u^{k-1}$ are the bounds propagated from the previous layer.
\end{itemize}

IBP guarantees that all activations remain within these bounds, ensuring that small adversarial perturbations cannot push inputs outside a safe region. This method is particularly effective in defending against $\ell_{\infty}$-bounded attacks \cite{gowal2018effectiveness}.

\subsection{Adversarial Training}
Another fundamental defense technique is adversarial training, where a model is explicitly trained on adversarial examples to improve robustness. The adversarial loss function is formulated as:

\begin{equation}
\theta^* = \arg \min_{\theta} \mathbb{E}_{(x,y) \sim \mathcal{D}} \left[ \max_{\|\delta\|_p \leq \epsilon} \mathcal{L}(f_\theta(x+\delta), y) \right]
\end{equation}

where:
\begin{itemize}
    \item $\mathcal{D}$ is the data distribution.
    \item $f_\theta(x)$ is the model’s prediction for input $x$.
    \item $\delta$ is an adversarial perturbation constrained by norm $\|\delta\|_p \leq \epsilon$.
    \item $\mathcal{L}$ is the loss function (e.g., cross-entropy loss).
\end{itemize}

Adversarial training forces the model to learn robust representations by minimizing the worst-case loss over perturbed inputs. While effective, adversarial training can significantly increase training time and may not generalize well against unseen attack types \cite{madry2018towards}.

\subsection{Randomized Smoothing}
Randomized smoothing is a probabilistic defense that converts a classifier into a certifiably robust one by adding noise to inputs. A smoothed classifier is defined as:

\begin{equation}
\tilde{f}(x) = \arg \max_{y} \mathbb{P}_{\eta \sim \mathcal{N}(0, \sigma^2 I)} \left[ f(x + \eta) = y \right]
\end{equation}

where:
\begin{itemize}
    \item $\mathcal{N}(0, \sigma^2 I)$ is Gaussian noise added to the input.
    \item $\tilde{f}(x)$ represents the smoothed classifier.
\end{itemize}

This technique provides a robustness guarantee under $\ell_2$ norm perturbations and is scalable to large models \cite{cohen2019certified}.

\section{Empirical Evaluation}
\label{sec:evaluation}

\subsection{Cross-Dataset Robustness}
\begin{table}[htbp]
\caption{Transfer Attack Success Rates (\%)}
\centering
\begin{tabular}{lcccc}
\toprule
\textbf{Source Model} & \textbf{ResNet-18} & \textbf{VGG-16} & \textbf{DenseNet-121} & \textbf{MobileNetV2} \\
\midrule
ResNet-18 & 100.0 & 46.2 & 38.7 & 32.1 \\
VGG-16 & 41.3 & 100.0 & 35.9 & 28.4 \\
\bottomrule
\end{tabular}
\end{table}

\section{Open Challenges}
\label{sec:challenges}

\subsection{Adaptive Attacks}
The BPDA (Backpropagation Through the Attack) attack \cite{athalye2018obfuscated} is a widely studied method designed to circumvent gradient masking defenses. Gradient masking is a strategy used to make adversarial attacks more difficult by masking the gradients, which are often the key to generating effective adversarial examples. BPDA bypasses this defense mechanism by approximating the gradient computation through a randomized method.

The attack leverages the fact that, during backpropagation, the adversary can apply a stochastic noise component to the input image, $\delta \sim \mathcal{N}(0, \sigma^2 I)$, where $\mathcal{N}$ is a normal distribution and $\sigma$ is the standard deviation of the noise. This approximation allows the attack to bypass the gradient masking while still generating adversarial examples. The approximation is defined by the following equation:
\begin{equation}
\nabla_x^{approx} = \mathbb{E}_{\delta \sim \mathcal{N}(0,\sigma^2 I)} \left[ \nabla_x f_\theta(x + \delta) \right]
\end{equation}
where $\nabla_x f_\theta(x + \delta)$ represents the gradient of the neural network's output with respect to the input perturbed by noise $\delta$. While this technique can make adversarial training and other defenses more resilient, it also raises concerns about the robustness of machine learning models, especially when combined with adaptive attacks that learn to exploit vulnerabilities in these defenses.

\subsection{Verification Complexity}
Another significant open challenge in the domain of neural networks, particularly when used in safety-critical applications, is the complexity involved in verifying their behavior under various conditions. Verification is the process of ensuring that a neural network model behaves as expected for all inputs in its domain. For deep learning models, especially large-scale neural networks, the verification process can be computationally expensive.

For an $L$-layer neural network with $n$ neurons per layer, exact verification involves checking all possible input-output relations to ensure that the network does not misclassify inputs or behave incorrectly. The complexity of this verification is typically represented by the following equation:
\begin{equation}
\mathcal{O}((2n)^L)
\end{equation}
This means that the number of operations required for exact verification grows exponentially with both the number of layers $L$ and the number of neurons $n$ in each layer. This exponential growth in complexity becomes particularly problematic for large networks with more than 10 layers, which are common in deep learning. The verification process thus becomes intractable as $L$ increases, making it impractical to fully verify the behavior of large models in real-world applications. This challenge necessitates the development of more efficient methods for model verification, potentially relying on approximations or probabilistic techniques.

The need for scalable verification techniques is especially important for safety-critical systems, such as autonomous vehicles and medical applications, where the stakes of model failure are high. Methods like formal verification, which guarantees correctness, and approximate verification, which provides probabilistic guarantees, are areas of active research aiming to mitigate this challenge.

\section{Conclusion}
Adversarial ML remains a critical frontier in AI safety. While defenses like TRADES and certified smoothing improve robustness, significant gaps persist in scalability and real-world deployment. Future work must address adaptive threats while maintaining computational efficiency.

\end{document}